\documentclass[
    ,final            
  ]
  {aipproc}

\layoutstyle{6x9}

\begin{document}

\title{$\Lambda(1520)$ at finite density}


\author{M. M. Kaskulov and E. Oset}{
  address={Departamento de F\'{\i}sica Te\'orica and IFIC,
Centro Mixto Universidad de Valencia-CSIC,
Institutos de
Investigaci\'on de Paterna, Aptd. 22085, 46071 Valencia, Spain}
}



\begin{abstract}
We study the decay channels of the $\Lambda(1520)$ 
in a nuclear medium
and find a sizable change - of the order of factor five - of the width of this
hyperon at normal nuclear matter density.
The mass shift of the $\Lambda(1520)$ is moderate.
%
%
\end{abstract}

\maketitle


  The change of resonance properties in the nuclear medium is a field that
captures permanent attention, and basic symmetries can be tested through medium
modification of particle properties. The later is 
particularly interesting for resonances which are not the genuine 
three quark states 
surviving in the large-$N_c$ limit of QCD, but can be generated dynamically
using the effective degrees of freedom and effective forces  
constrained by the symmetry pattern underlying
the QCD.

The $\Lambda(1520)$ is an example of such dynamic state which 
can be generated from the $s$-wave interaction of the decuplet of baryons
with the octet of pseudoscalar mesons
using the leading order chiral Lagrangian 
in combination with resummation schemes which 
respect the unitarity in coupled channels~\cite{kolo,sarkar}. 
In particular, 
the  $\Lambda(1520)$ appears basically as a quasibound state of the 
$\pi \Sigma^*(1385)$ system. 

The small free width of the 
$\Lambda(1520)$ of $\simeq 15.6$~MeV comes from the decay into
$\bar{K}N$
and $\pi \Sigma$ channels, but the decay into 
$\pi \Sigma^*(1385)$ is forbidden 
for the nominal mass of the $\Sigma^*(1385)$.  The coupling of the 
$\Lambda(1520)$ to $\bar{K}N$ and $\pi \Sigma$ 
makes the picture of the $\Lambda(1520)$ more elaborate, with 
$\pi \Sigma^*(1385)$ being a very important component but with also
sizable admixtures of $\bar{K}N$ and $\pi \Sigma$~\cite{Sarkar:2005sp}.

At finite baryonic density, 
the decay of the $\Lambda(1520)$ 
bears resemblance to the one of the $\Delta(1232)$ in the nuclear medium. 
The  $\Delta$ decays into
$\pi N$ and the $\pi$ gets renormalized in the medium by exciting $ph$ and
$\Delta h$ components, as a consequence of which the $\Delta$ is renormalized
and its pion (photon) induced excitation in nuclei incorporates now the
mechanisms of pion (photon)
absorption in the medium. In the present case, the $\Lambda(1520)$ decay into
$\pi \Sigma^*(1385)$, only allowed through the $\Sigma^*(1385)$ width, gets
drastically
modified when the $\pi$ is allowed to excite $ph$ and $\Delta h$ components in
the nucleus, since automatically the phase space for the decay into $ph
\Sigma^*(1385)$ gets tremendously increased. This fact, together with the
large coupling of the $\Lambda(1520)$ to the $\pi \Sigma^*(1385)$ channel
predicted by the chiral unitary approach, leads to a very large width of the
$\Lambda(1520)$ in nuclei. Similar nuclear effects will
modify the $\pi \Sigma$ decay channel and the $\bar{K}N$ will be analogously
modified when the $\bar{K}$ is allowed to excite hyperon-hole excitations. All
these channels lead to a considerable increase of the width of the 
$\Lambda(1520)$ in the nucleus.


In the description of the $\Lambda(1520)$  properties in the  nuclear medium
we closely follow the formalism developed in Ref.~\cite{murat} (see also
Ref.~\cite{Lutz:2001dq} for related discussions and 
comments in Ref.~\cite{murat} about 
the approximation made in Ref.~\cite{Lutz:2001dq}).
Here we briefly summarize the main results of that studies.

In the nuclear medium the $\Lambda(1520)$ gets renormalized through
the conventional 
$d$-wave decay channels including $\Lambda(1520) \to \bar{K}N$ and
$\Lambda(1520) \to \pi \Sigma$ which 
accounts for practically all of the $\Lambda(1520)$ free 
width. 
In addition in Ref.~\cite{murat} 
as a novel element the $s$-wave decay $\Lambda(1520) \to \pi \Sigma^*(1385)$ 
has been considered which is forbidden in the free space 
for the nominal masses of the $\Lambda(1520)$ and $\Sigma^*(1385)$
but opens in the nuclear medium.


\begin{figure}
  \includegraphics[height=.3\textheight]{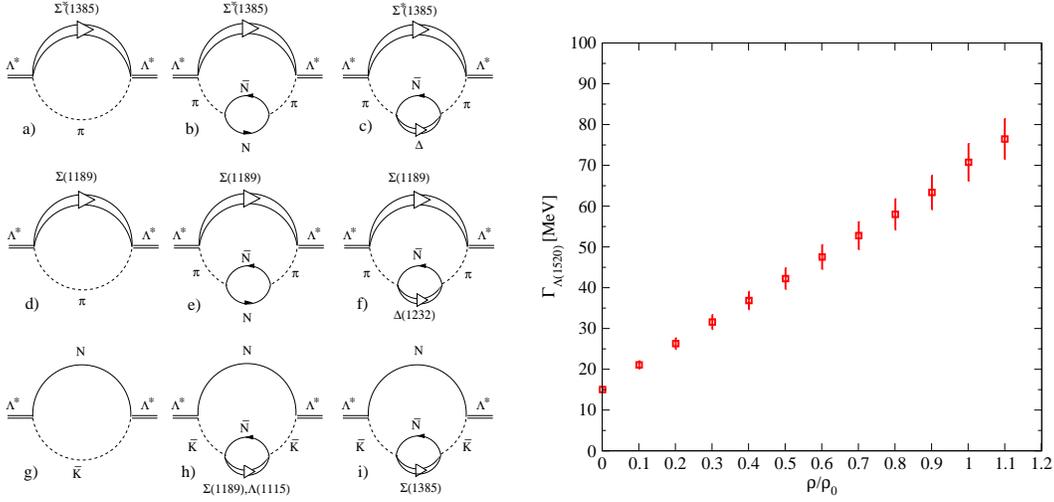} \hspace{0.4cm}
  \includegraphics[height=.28\textheight]{Fig2.eps}
  \caption{\label{F1}Left panel: The renormalization of the 
$\Lambda(1520)$ in the nuclear medium. Right panel: 
Values with theoretical uncertainties for the width of the $\Lambda(1520)$
at rest in the medium, including the free width, 
as function of the nuclear matter density
$\rho/\rho_0$ (a).}
\end{figure}

The model diagrams which describe the renormalization
of the $\Lambda(1520)$ in the nuclear medium are shown in Fig.~\ref{F1}
(left panel).
The in-medium propagation of pions in the loops is affected by the 
excitation of the $p-hole$ and $\Delta(1232)-hole$ states and in the antikaon
$\bar{K}$ case by the excitation of the all relevant hyperon-hole states. 
The intermediate baryons in the loops  are also dressed  with respect
to their own decay channels properly renormalized in the nuclear medium. 
The later includes the dressing by means of the
phenomenological optical potentials which account for the nuclear binding
corrections, Pauli blocking for the nucleons and
short-range correlations in the $p$-wave transitions induced
by the strong repulsive forces at short inter-baryon 
distances of the Landau-Migdal type. 
In Fig.~\ref{F1} (right panel) we show the model prediction 
for the width $\Gamma_{\Lambda^*}$ 
of the $\Lambda(1520)$ at rest and at the nominal pole position
as a function of the nuclear matter density $\rho/\rho_0$ where
$\rho_0=0.16$~fm$^{-3}$ is the normal nuclear density. The error bar reflects
the theoretical uncertainties due to the choice of the momentum 
cut off in the $d$-wave loops.
As one can see the model predicts a spectacular change of the
width of the $\Lambda(1520)$ in the nuclear medium
which gets increased by factor
$\sim 5$ at normal nuclear matter densities. The corresponding mass
shift of the $\Lambda(1520)$ is $\delta m \simeq - 30$~MeV only.

\begin{figure}
  \includegraphics[height=.15\textheight]{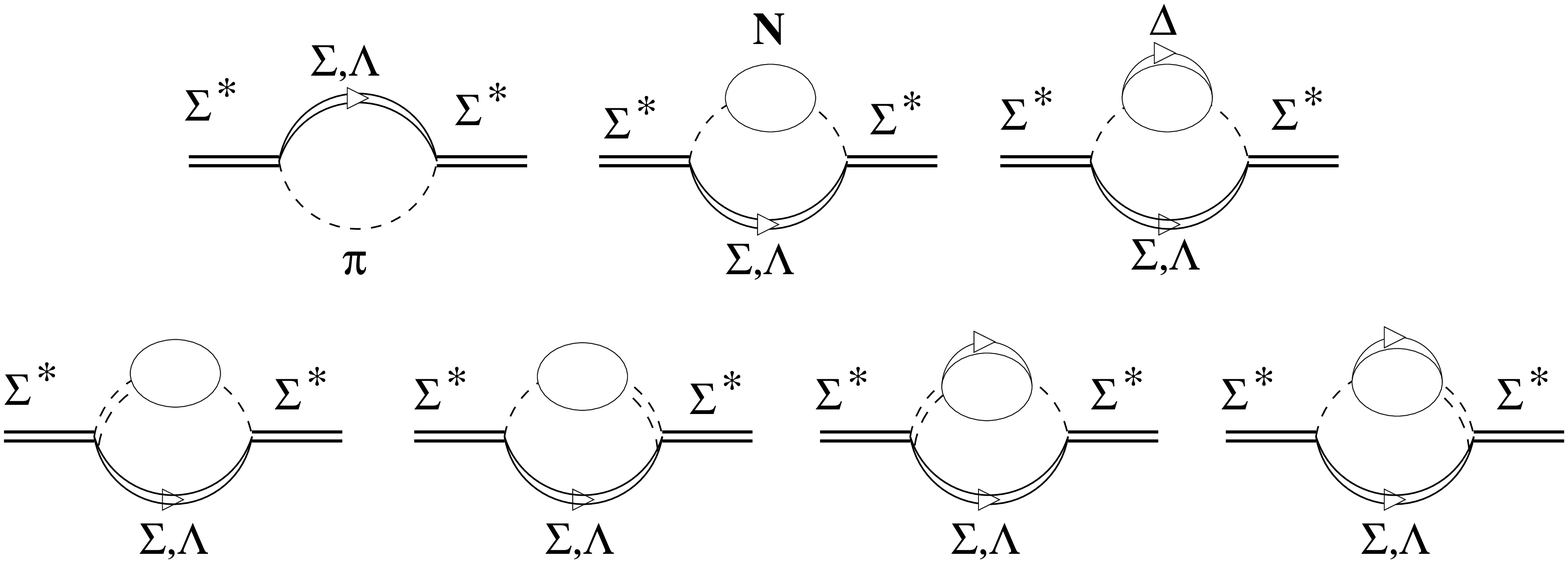}
  \caption{In-medium renormalization of the $\Sigma^*(1385)$ in the 
$\pi \Sigma + \pi \Lambda$ channels. The last four
diagrams account for the short-range correlations. \label{F2}}
\end{figure}

The problem of the in-medium modification of the $\Lambda(1520)$ 
in the $\pi \Sigma^*(1385)$ channel is mainly
reduced to the proper description of the properties of pions 
and $\Sigma^*(1385)$ at finite nuclear density.
The renormalization scheme which we
employ for the $\Sigma^*(1385)$ is essentially
the same as in the previous case except for the $p$-wave nature of 
the hadronic $\Sigma^*(1385)$ decay. This implies some peculiarities, for
instance, the proper treatment of short range correlations.
The typical diagrams describing the in-medium selfenergy of 
$\Sigma^*(1385)$ are shown in Fig.~\ref{F2}. The later account for
the $\Sigma^*(1385) \to \pi \Sigma + \pi \Lambda$ decay channels. Similar
diagrams appear then considering the $\bar{K}N$ decay channel where
$\bar{K}$ gets renormalized 
by coupling to the all relevant hyperon-hole excitations. 
All together,  we find an increase of the width of the
$\Sigma^*(1385)$ by the factor $\sim 2\div 3$ at normal nuclear matter
density
with respect to the free width $\Gamma_{free} \simeq 35$~MeV. The in-medium
change of the mass of the $\Sigma(1385)$ is very moderate $\delta m \simeq
-40$~MeV.

The spectacular change in the $\Lambda(1520)$ width should be easily verified
experimentally. To facilitate this work we have performed calculations of the 
$A$ dependence of the $\Lambda(1520)$ produced in the $\gamma A \to K^+ 
\Lambda(1520) A'$ and $p A \to p'K^+ \Lambda(1520) A'$ reactions where we find
a sizable $A$ dependence linked to the modification of the $\Lambda(1520)$
width
in the medium~\cite{KRO}.



{\bf Acknowledgments}:
This work is partly supported by DGICYT contract number BFM2003-00856,
and the E.U. EURIDICE network contract no. HPRN-CT-2002-00311.
This research is part of the EU Integrated Infrastructure Initiative
Hadron Physics Project under contract number RII3-CT-2004-506078.



\bibliographystyle{aipproc}   

\bibliography{sample}

\begin{thebibliography}{9}



\bibitem{kolo}
  E.~E.~Kolomeitsev and M.~F.~M.~Lutz,
  Phys.\ Lett.\ B {\bf 585} (2004) 243.

\bibitem{sarkar}
  S.~Sarkar, E.~Oset and M.~J.~Vicente Vacas,
  Nucl.\ Phys.\ A {\bf 750} (2005) 294.

\bibitem{Sarkar:2005sp}
  S.~Sarkar, L.~Roca, E.~Oset, V.~K.~Magas and M.~J.~V.~Vacas,
  arXiv:nucl-th/0511062.

\bibitem{murat}
M.~Kaskulov and E.~Oset,
arXiv:nucl-th/0509088.

\bibitem{Lutz:2001dq}
  M.~F.~M.~Lutz and C.~L.~Korpa,
  Nucl.\ Phys.\ A {\bf 700}, 309 (2002)

\bibitem{KRO}
M. Kaskulov, L. Roca and E. Oset, in preparation.
\end{thebibliography}

\IfFileExists{\jobname.bbl}{}
 {\typeout{}
  \typeout{******************************************}
  \typeout{** Please run "bibtex \jobname" to optain}
  \typeout{** the bibliography and then re-run LaTeX}
  \typeout{** twice to fix the references!}
  \typeout{******************************************}
  \typeout{}
 }


\end{document}